\documentclass[fleqn,twoside]{article}
\usepackage{espcrc2}
\usepackage{graphicx}
\usepackage{epsfig}

\def\gtap{\mathrel{ \rlap{\raise 0.511ex \hbox{$>$}}{\lower 0.511ex
   \hbox{$\sim$}}}} 
\def\ltap{\mathrel{ \rlap{\raise 0.511ex
   \hbox{$<$}}{\lower 0.511ex \hbox{$\sim$}}}}

\title{Atmospheric neutrinos and $\nu$-mass hierarchy}

\author{Sergio Palomares-Ruiz\address[UV]{Departamento de F\'{\i}sica 
      Te\'orica and IFIC, Universidad de Valencia-CSIC, 46100
      Burjassot, Valencia, Spain}\address{Department of Physics and
      Astronomy, UCLA, Los Angeles, CA 90095, USA}\address{Department
      of Physics and Astronomy, Vanderbilt University, Nashville TN
      37235, USA}\thanks{Talk presented at TAUP 2003, Seattle, USA,
      5-9 September 2003.}, Jos\'e Bernab\'eu\addressmark[UV]
      }

\begin{document}

\begin{abstract}
We discuss the possibility for matter effects in the
three-neutrino oscillations of the atmospheric $\nu_e$ ($\bar{\nu}_e$)
and $\nu_\mu$ ($\bar{\nu}_\mu$), driven by one neutrino mass squared
difference, $|\Delta m^2_{31}| \gg \Delta m^2_{21}$, to be observable
under appropriate conditions. We derive predictions for the Nadir
angle ($\theta_n$) dependence of the ratio $N_{\mu}/N_e$ 
of the rates of the $\mu-$like and $e-$like multi-GeV events which is
particularly sensitive to the Earth matter effects in the  
atmospheric neutrino oscillations, and thus to the values of
$\sin^2\theta_{13}$ and $\sin^2\theta_{23}$, and also   
to the type of neutrino mass spectrum.
\end{abstract}

\maketitle

\section{Introduction}
Present evidence for neutrino masses and mixings can be summarized as:
1) the atmospheric $|\Delta m^{2}_{31}|$ is associated with a mixing
$\theta_{23}$ close to maximal \cite{SKatm00}; 2) the solar $\Delta
m^{2}_{21}$ prefers the LMA-MSW solution \cite{solar,KamLAND}; 3)
CHOOZ reactor data \cite{CHOOZ} give severe limits for $|U_{e
  3}|$. Here, we discuss that contrary to a wide spread belief, Earth
effects on the propagation of atmospheric neutrinos can become
observable, in detectors with lepton charge-discrimination
\cite{mantle,core} and in water-\v{C}erenkov ones
\cite{th13cerenkov}, even if $|U_{e 3}|$ is small, yet
non-vanishing. This fact could allow to determine the sign of $\Delta
m^{2}_{31}$ and obtain more stringent constraints on the values of
$\theta_{13}$ and $\theta_{23}$. Getting more precise information
about the value of these mixing angles and determining the type of the
neutrino mass spectrum (with normal or inverted hierarchy) with a
higher precision is of fundamental importance for the progress in the
studies of neutrino mixing.   

We study here the possibilities to obtain this type of information
using the atmospheric neutrino data that can be provided by present
and future water-\v{C}erenkov detectors. For baselines $L$ smaller
than the Earth diameter, appropiate for atmospheric neutrinos,
$\frac{\Delta m^{2}_{21}}{4 E} L \equiv \Delta_{21} \ll 1$, so that we
will neglect the (1,2)-oscillating phase in vacuum against the
(2,3)-one. This is a very good aproximation for the presently best
favored LMA-I solution to the solar neutrino problem \cite{solar}.   

The Earth matter effects, which can resonantly enhance the $\nu_{\mu}
\rightarrow \nu_e$ and $\nu_{e} \rightarrow \nu_{\mu(\tau)}$
transitions, lead to the reduction of the rate of the multi-GeV
$\mu-$like events and to the increase of the rate of the multi-GeV
$e-$like events in a water-\v{C}erenkov detector with respect to the
case of absence of these transitions (see, e.g.,
\cite{core,SP3198}). Correspondingly, as observables which are 
sensitive to the Earth matter effects, we consider the Nadir-angle
distribution of the ratio $N_{\mu}/N_{e}$, where $N_{\mu}$ and $N_e$
are the multi-GeV $\mu-$like and  $e$-like numbers of events,
respectively.

\section{The matter-induced neutrino spectrum}
If $V$ is the effective neutrino potential, in going from $\nu$ to
$\overline{\nu}$, there are matter-induced CP- and CPT- odd effects
associated with the change $V \rightarrow - V$. The effects here
discussed depend on the interference between the different flavors and
on the relative sign between $2 E V$ and $\Delta m^{2}_{31}$. For
atmospheric neutrinos, an appreciable interference will be present if
and only if there are appreciable matter effects, i.e., the
``connecting'' mixing $U_{e 3}$ between the $\nu_e$-flavor and the
$\nu_3$ mass eigenstate does not vanish.     

For $s_{13} \equiv \sin{\theta_{13}} = 0$, matter effects lead to a
breaking of the (1,2)-degeneracy such that $\tilde{\nu}_2$ coincides
with $\nu_e$. The net effect is that $\nu^{m}_{1}$ and $\nu^{m}_{3}$
lead to the atmospheric $\nu_{\mu} \rightarrow \nu_{\tau}$ indicated
by SK, likewise in vacuum the (2,3)-mixing does. The $\nu_e$-flavor
decouples in matter, even if there was a large mixing in the
(1,2)-system. No matter effects would then be expected when starting
with $\nu_{\mu}$, i.e., there would be no chances to distinguish the
type of mass hierarchy by these means. However, for small $s_{13}$,
even if the effects on the spectrum are expected to be small, there
could be a substantial mixing of $\nu_e$ with $\nu^{m}_{3}$. This
would lead to a resonant MSW behaviour in the case of neutrinos
crossing only the Earth mantle and to new resonant effects (NOLR)
\cite{SP3198} in the case of neutrinos crossing also the Earth
core. But still $\langle \nu^{m}_{1} | \nu_e \rangle = 0$. This
vanishing mixing in matter is responsible for the absence of
fundamental CP-violating effects, even if there are three
non-degenerate mass eigenstates in matter. In vacuum, the absence of
genuine CP-odd probabilities was due to the degeneracy $\Delta_{21} =
0$.

\section{Atmospheric neutrino oscillations in the Earth}

The fluxes of atmospheric $\nu_{e,\mu}$ of energy $E$, which reach the
detector after crossing the Earth along a given trajectory  specified
by the value of $\theta_{n}$, $\Phi_{\nu_{e,\mu}}(E,\theta_{n})$, 
are given by the following expressions in the case of the
three-neutrino oscillations under discussion \cite{SP3198}: 

\begin{equation}
\Phi_{\nu_e}(E,\theta_{n}) \cong 
\Phi^{0}_{\nu_e}~\left (  1 + 
  [s^2_{23}r - 1]~P_{2\nu}\right )
\label{Phie}
\end{equation}
\vspace{-3mm}
\begin{eqnarray}
\label{Phimu}
\Phi_{\nu_{\mu}}(E,\theta_{n}) \hspace{-3mm} & \cong \hspace{-2.5mm} &
\Phi^{0}_{\nu_{\mu}} \left( 1 + s^4_{23}~ [(s^2_{23}~r)^{-1} -
  1]~P_{2\nu} \right. \nonumber \\ 
  & & \left. - 2c^2_{23}s^2_{23}~\left [ 1 - Re~( e^{-i\kappa}
  A_{2\nu}) \right ] \right)   
\end{eqnarray}

\noindent 
where $P_{2\nu} \equiv P_{2\nu}(\Delta m^2_{31},
\theta_{13};E,\theta_{n})$ is the probability of two-neutrino
oscillations in the Earth, $\kappa$ and $A_{2\nu}$ are known phase and
two-neutrino transition probability amplitude,
$\Phi^{0}_{\nu_{e(\mu)}}$ is the $\nu_{e(\mu)}$ flux in the absence of
neutrino oscillations and  

\begin{equation}
r \equiv r(E,\theta_{n}) \equiv
\frac{\Phi^{0}_{\nu_{\mu}}(E,\theta_{n})} 
{\Phi^{0}_{\nu_{e}}(E,\theta_{n})}~~.
\label{r}
\end{equation}

The predicted ratio \cite{flux} for atmospheric sub-GeV neutrinos is
$r \cong (2.0 - 2.5)$, whereas $r \cong (2.6 - 4.5)$ for multi-GeV
atmospheric neutrinos. If $s^2_{23} = 0.5$, the possible effects of
the  $\nu_{\mu} \rightarrow \nu_{e}$ and $\nu_{e} \rightarrow \nu_{\mu
  (\tau)}$ transitions on the sub-GeV $e-$like events would be
strongly suppressed even if these transitions were maximally enhanced
by the Earth matter effects. On the other hand, $r > 2$ for the
multi-GeV sample, and matter effects can show up. Thus, in the case
under study, the effects of the $\nu_{\mu} \rightarrow \nu_{e}$
and $\nu_{e} \rightarrow \nu_{\mu (\tau)}$ oscillations, increase
with the increase of $s^2_{23}$, are considerably larger in the
multi-GeV samples of events than in the sub-GeV ones and in the
multi-GeV case, they lead to an increase of the rate of
$e-$like events and to a slight decrease of the $\mu-$like event
rate. This discussion suggests that the quantity most sensitive to the
effects of the oscillations of interest should be the ratio of the
$\mu-$like and $e-$like multi-GeV events, $N_{\mu}/N_{e}$.

If $s_{13} \ne 0$, the Earth matter effects can resonantly enhance
$P_{2\nu}$ for $\Delta m^2_{31} > 0$ and $\bar{P}_{2\nu}$ if $\Delta
m^2_{31} < 0$ \cite{mantle}. Due to the difference of cross sections 
for neutrinos and antineutrinos, approximately 2/3 of the total rate 
of the $\mu-$like and $e-$like multi-GeV atmospheric neutrino 
events in a water-\v{C}erenkov detector, i.e., $\sim 2N_{\mu}/3$ and
$\sim 2N_e/3$, are due to neutrinos $\nu_{\mu}$ and $\nu_e$,
respectively, while the remaining $\sim 1/3$ of the multi-GeV  event
rates, i.e., $\sim N_{\mu}/3$ and $\sim N_e/3$, are produced by
antineutrinos $\bar{\nu}_{\mu}$ and $\bar{\nu}_e$. This implies that
the Earth matter effects in the multi-GeV samples of $\mu-$like and
$e-$like events will be larger if $\Delta m^2_{31} > 0$ (normal
hierarchy), than if $\Delta m^2_{31} < 0$ (inverted hierarchy). Thus,
the ratio $N_{\mu}/N_e$ of the multi-GeV $\mu-$like and $e-$like event
rates measured in water-\v{C}erenkov detectors is sensitive, in
principle, to the type of the neutrino mass spectrum
\cite{th13cerenkov}.

\section{Results}

In Fig. 1 we show the predicted dependences on $\cos\theta_n$ of the
ratios of the multi-GeV $\mu-$ and $e-$ like events, integrated over
the neutrino energy from the interval $E = (2 - 10)$ GeV, for
different cases (see figure). 

For $\cos\theta_n \ltap 0.2$ and $|\Delta m^2_{31}| = 3\times
10^{-3}~{\rm eV^2}$, the oscillations of the atmospheric 
$\nu_e$ and $\bar{\nu}_e$ with energies in the multi-GeV range $E \sim
(2 - 10)$ GeV, are suppressed. If $\Delta m^2_{31} > 0$, for instance,
the Earth matter effects suppress the antineutrino 

\vspace{-5mm}
\centerline{\epsfysize=8.5cm \epsfbox{ratiomue.eps}}
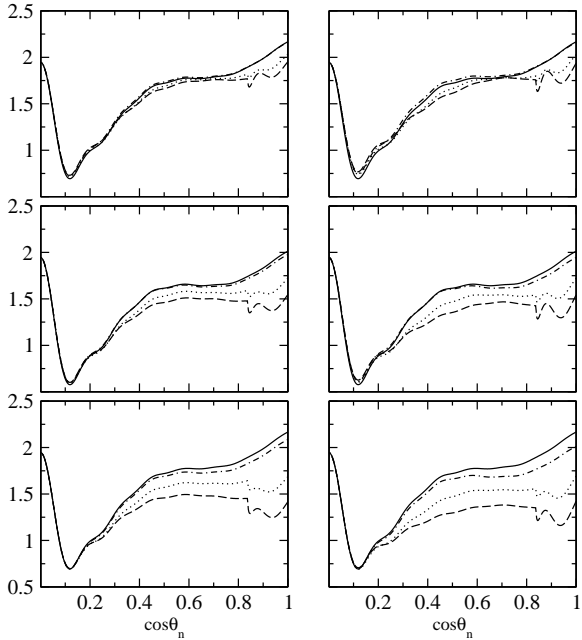
 { Figure 1. The dependence on $\cos\theta_n$ of the ratios of the
 multi-GeV $\mu-$ and $e-$ like events, integrated over the 
 neutrino energy in the interval $E = (2 - 10)$ GeV, in the cases  
i) of two-neutrino $\nu_{\mu} \rightarrow \nu_{\tau}$ oscillations in
 vacuum and no $\nu_e$ oscillations, $N^{2\nu}_{\mu}/N^{0}_{e}$ (solid
 lines), ii) three-neutrino oscillations in vacuum of $\nu_{\mu}$
  and $\nu_e$, $(N^{3\nu}_{\mu}/N^{3\nu}_{e})_{vac}$ (dash-dotted
 lines), iii) three-neutrino oscillations of $\nu_{\mu}$ and $\nu_e$
 and in the Earth and neutrino mass spectrum with normal hierarchy
 $(N^{3\nu}_{\mu}/N^{3\nu}_{e})_{\rm NH}$ (dashed lines), or with
 inverted hierarchy, $(N^{3\nu}_{\mu}/N^{3\nu}_{e})_{\rm IH}$ (dotted
 lines). The results shown are for $|\Delta m^2_{31}| = 3\times
 10^{-3}~{\rm eV^2}$, $\sin^2\theta_{23} = 0.36~(\rm upper~panels);~0.50
~(\rm middle~panels);~0.64$ (lower panels), and
 $\sin^22\theta_{13} = 0.05$ (left panels); $0.10 (\rm
 right~panels)$.}
\vspace{3mm}

\noindent
oscillation probability $\bar{P}_{2\nu}$, but can enhance the neutrino
mixing in matter. However, since the neutrino path in the Earth mantle
is relatively short, $P_{2\nu} \ll 1$.

At $\cos\theta_n \gtap 0.4$, the Earth matter effects in the
oscillations of the atmospheric $\nu_{\mu}$ and $\nu_e$ can generate
noticeable differences between $N^{2\nu}_{\mu}/N^{0}_{e}$ (or
$(N^{3\nu}_{\mu}/N^{3\nu}_{e})_{vac}$) and
$(N^{3\nu}_{\mu}/N^{3\nu}_{e})_{\rm NH(IH)}$, as well as between
$(N^{3\nu}_{\mu}/N^{3\nu}_{e})_{\rm NH}$ and
$(N^{3\nu}_{\mu}/N^{3\nu}_{e})_{\rm IH}$. 

\section{Conclusions}
We have studied the possibility to obtain evidences for Earth matter
enhanced atmospheric neutrino oscillations involving, in particular,
the $\nu_e$, from the analysis of the $\mu-$like and $e-$like
multi-GeV event data that can be provided by present and future
water-\v{C}erenkov detectors. We have seen that such evidences could
give also important quantitative information on the values of
$\sin^2\theta_{13}$ and $\sin^2\theta_{23}$ and on the sign of $\Delta
m^2_{31}$. 
\vspace{2mm}

{\bf Acknowledgments} --- This work is supported by the Spanish Grant
FPA2002-00612 of the MCT, by the Spanish MCD and SPR in part by NASA
Grant ATP02-0000-0151.

\end{document}